\documentclass[12pt]{iopart}
\usepackage{iopams}
\usepackage{graphicx}
\usepackage{bm}
\usepackage{color}
\usepackage{float}
\usepackage{dcolumn}
\usepackage{cite}
\usepackage{enumerate}
\usepackage{multirow}
\usepackage{threeparttable}
\usepackage{hyperref}
\hypersetup{
  colorlinks= true,
  linkcolor=blue,anchorcolor=blue,citecolor=blue,urlcolor=blue
  }
\expandafter\let\csname equation*\endcsname\relax
\expandafter\let\csname endequation*\endcsname\relax
\usepackage{amsmath,amssymb}

\begin{document}

\title{Extending pairing energy density functional using pairing rotational moments of inertia}

\author{Nobuo Hinohara}

\address{
  Center for Computational Sciences, University of Tsukuba, Tsukuba, 305-8577, Japan}
\ead{hinohara@ccs.tsukuba.ac.jp}

\vspace{10pt}
\begin{indented}
\item[] 
\end{indented}

\begin{abstract}
  The pairing energy density functionals (EDFs) that include the spatial derivative and kinetic terms of the pair densities are discussed. 
  The coupling constants of the pairing EDF are adjusted to reproduce the experimental
  pairing rotational moment of inertia, and the pair-density derivative terms are shown to systematically improve
  the values of the pairing rotational moments of inertia in Sn and Pb isotopes.
  It is pointed out that the 
  conventional average pairing gaps  overestimate the experimental odd-even mass staggering in the presence of the pair-density derivative terms.
\end{abstract}

\vspace{2pc}
\noindent{\it Keywords}: spontaneous symmetry breaking, pairing rotation, QRPA, energy density functional
%\submitto{\JPG}
%\maketitle

\section{Introduction}

Pairing is one of the important many-body correlations in
finite nuclei and other quantum systems \cite{RevModPhys.75.607,Brink-Broglia,50yearsBCS}.
The existence of the pairing-type correlation in finite nuclei has been known by a number of experimental data.
The odd-even staggering of the binding energy (OES) is the most commonly used observable relevant to the nuclear pairing.
In systems with even number of neutrons (or protons), all the neutrons can form pairs and thus obtain maximal pairing correlations,
while in systems with odd number of neutrons, one last neutron cannot form a pair, and the systems are less bound than the neighboring systems with even number of neutrons.

Pairing correlations can be taken into account within the mean-field approximations such as 
Bardeen-Cooper-Schrieffer (BCS) and Hartree-Fock-Bogoliubov (HFB) theories \cite{Ring-Schuck}.
Quasiparticles which are linear combination of the particle creation and annihilation
allow us to describe a superconducting ground state as a one-body field.
For the qualitative and quantitative description of the pairing properties of the atomic nuclei,
it is essential to identify appropriate observable that is sensitive to the
detail of the nuclear pairing.
Conventionally the theoretical pairing gap is compared with the experimental OES,
since the direct computation of the odd-mass systems in the mean-field approximation is not as accurate as that of the even-mass systems.
As the relation between the pairing gap and OES is indirect,
it is not easy to discuss the detailed information on the pairing correlation from the pairing gap without ambiguity.

In the mean-field approximation based on a nuclear effective interaction,
the particle-hole and pairing potentials are derived from a single nuclear effective Hamiltonian.
In actual applications based on the zero-range Skyrme effective interactions
(with a few exceptions such as SkP \cite{Dobaczewski1984103} and application to the nuclear matter \cite{Takahara1994261,ActaPhysPolBSuppl8_651}), however, Skyrme effective interactions are used only for the particle-hole channel, and simplified forms are assumed for the pairing interaction.
It is because that the pairing properties derived from the Skyrme interactions are found to be unrealistic \cite{RevModPhys.75.121}.
Such a current standard prescription for the mean-field theory with the Skyrme interactions is closely connected to the nuclear density functional theory (DFT). In nuclear DFT, the correspondence between the energy density functional (EDF) and the effective interaction is not required, and
the particle-hole and pairing parts of the EDF can be independent.

The coupling constants in the nuclear EDF is phenomenologically optimized using a selected experimental data set. Recently new EDF parametrizations such as
SV$_{\rm min}$ \cite{PhysRevC.79.034310} and 
UNEDF \cite{PhysRevC.82.024313,PhysRevC.85.024304,PhysRevC.89.054314,0954-3899-42-3-034024}, based on the optimizations using a large number of
experimental data set have been proposed.
Although many coupling constants are considered in the particle-hole channel,
the pairing functionals used there are still limited to a simple form.
This is because the optimization of the pairing part relies mainly on the theoretical pairing gap and experimental OES.
As commented in \cite{PhysRevC.82.024313}, a proper treatment of odd-mass systems including the evaluation of the time-odd fields is unavoidable in the optimization of the pairing functional in a more general form using the OES.

The pairing correlation introduces spontaneous breaking of the gauge symmetry.
A zero-energy Nambu-Goldstone (NG) mode \cite{PhysRev.112.1900,PhysRev.117.648,INC_19_154},
called pairing rotation appears
associated with the broken gauge symmetry \cite{Broglia20001}.
In our previous work \cite{PhysRevLett.116.152502}, we have proposed that the moments of inertia for the pairing rotation
that are related to the two-nucleon shell gap indicators $\delta_{2n}$ and $\delta_{2p}$ \cite{RevModPhys.75.1021}
and the proton-neutron interaction energy $\delta V_{pn}$ \cite{Zhang19891}, 
are excellent pairing indicators.
By employing the pairing rotational moments of inertia as pairing observables,
the issues related to the evaluation of the odd-mass systems can be avoided,
and the details of the pairing part of the nuclear EDF is expected to be constrained.

The aim of this article is to show the usefulness of the pairing rotational moments of inertia as nuclear pairing observables, and to demonstrate its sensitivity to the form of the pairing EDF.
We extend the pairing EDF by including the spacial derivative terms of the pair density. These terms are originated from
the momentum-dependent terms in the Skyrme effective interaction.
In section~\ref{sec:pairrot}, we describe the problem of using the conventional pairing gaps as approximations to the experimental OES.
Then we introduce the pairing rotational moment of inertia as a pairing observable.
In section~\ref{sec:FAM}, an efficient microscopic method for computing the pairing rotational moment of inertia
within the linear response theory for nuclear DFT is recapitulated \cite{PhysRevC.92.034321,PhysRevLett.116.152502}.
The extended form of the pairing EDF including the pair-density derivative terms
is presented in section \ref{sec:ppEDF},
and the results of the numerical calculations for neutron pairing in Sn and Pb isotopes are discussed in section \ref{sec:calc}.
Conclusions are given in section \ref{sec:conclusions}.

\section{Pairing observable \label{sec:pairrot}}

\subsection{Pairing gap and OES}

Conventionally the OES is used for the observable related to the nuclear pairing property \cite{RevModPhys.75.121}.
The simplest three-point formula for the OES is defined by the ground-state energy differences of the three neighboring isotopes \cite{PhysRevC.79.034306,PhysRevLett.81.3599}
\begin{align}
  \Delta^{(3)}_n(N) = \frac{(-1)^N}{2}[ E(N+1) - 2 E(N) + E(N-1)], \label{eq:OES}
\end{align}
where $E(N)$ is the ground-state energy of an isotope with $N$ neutrons.
In order to reduce the local fluctuation in the OES,
the average of the two OES values of even-odd and odd-even nuclei are used \cite{PhysRevC.89.054314}
\begin{align}
  \tilde{\Delta}_n^{(3)}(N) = \frac{1}{2}[ \Delta_n^{(3)}(N-1) + \Delta_n^{(3)}(N+1)].  \label{eq:OES2}
\end{align}
In this paper we use definition (\ref{eq:OES2}) for the experimental OES.

In the mean-field calculation, instead of evaluating the OES by either equation ~(\ref{eq:OES}) or (\ref{eq:OES2}), the theoretical pairing gap is usually
evaluated and compared with the OES.
For the conventional pairing EDF $\tilde{\chi}_n[\rho,\tilde{\rho},\tilde{\rho}^\ast] = \tilde{C}_n^{\rho}[\rho_0]|\tilde{\rho}_n(\bm{r})|^2$
with an isoscalar-density-dependent coupling constant $\tilde{C}_n^{\rho}[\rho_0]$,
the pairing gaps are defined by averaging the pairing potential $\tilde{h}_n(\bm{r})=\partial\tilde{\chi}_n/\partial\tilde{\rho}_n^\ast$ with the particle-hole or pair density. One definition of the neutron pairing gap is 
\begin{align}  
  \Delta_n(\rho) \equiv \frac{{\rm Tr}\tilde{h}_n\rho_n}{{\rm Tr}\rho_n}
  = -\frac{2}{N} \int {\rm d}\bm{r}  
  \tilde{C}_n^\rho[\rho_0] \rho_n(\bm{r}) \tilde{\rho}_n(\bm{r}), \label{eq:gaprho}
\end{align}
where $N$ is the neutron number. This is the pairing potential averaged with the particle-hole density, and is often used as the theoretical pairing gap \cite{Dobaczewski1984103,Bennaceur200596}.
Another definition of the pairing gap is the pairing potential averaged with the pair density and thus the focus is put more on the properties around the Fermi surface \cite{SauvageLetessier1981231,Bender2000}
\begin{align}
  \Delta_n(\tilde{\rho}) \equiv  \frac{{\rm Tr}\tilde{h}_n\tilde{\rho}_n}{{\rm Tr}\tilde{\rho}_n}
  = -\frac{2}{\tilde{N}} \int {\rm d}\bm{r}
   \tilde{C}_n^\rho[\rho_0] \tilde{\rho}_n^2(\bm{r}), \label{eq:gapkap}
\end{align}
where $\tilde{N}=\int {\rm d}\bm{r} \tilde{\rho}_n(\bm{r})$.

\begin{figure}[t]
  \begin{center}
    \includegraphics[width=150mm]{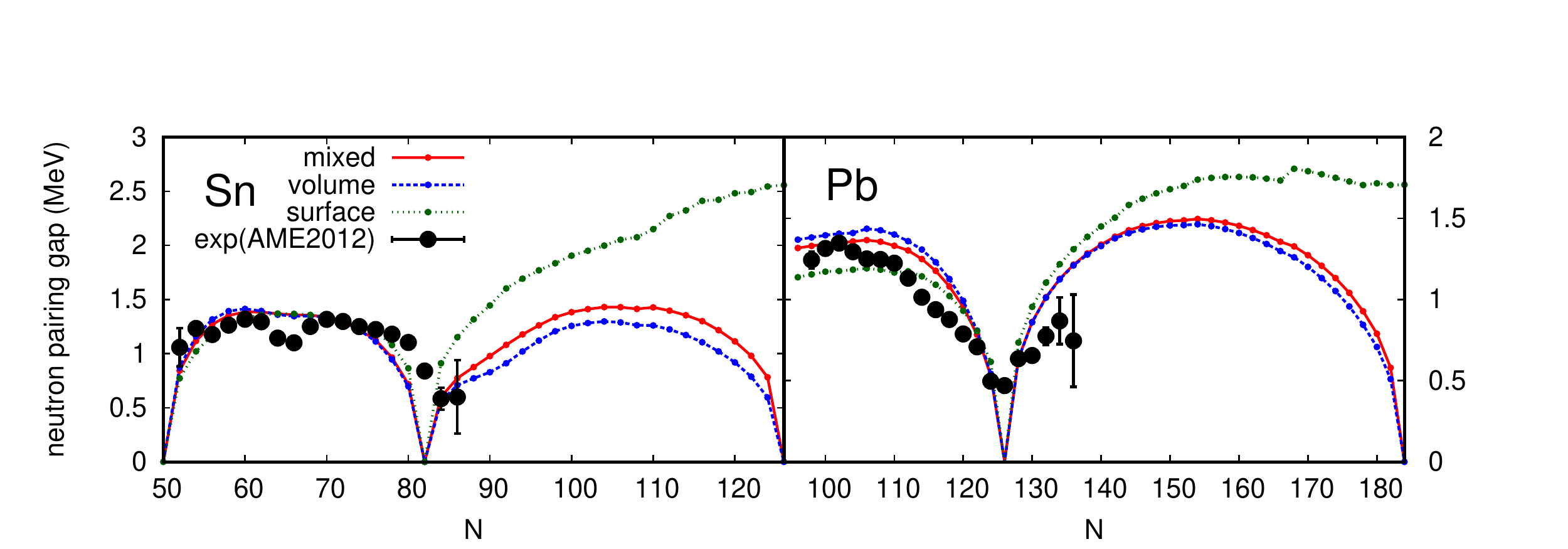}
    \end{center}
  \caption{
    Average neutron pairing gaps $\Delta_n(\rho)$ computed using the UNEDF1-HFB functional \cite{0954-3899-42-3-034024} with three kinds of density-dependent pairing functionals
    and the experimental OES evaluated with equation (\ref{eq:OES2}) using the values in \cite{ame2012} for Sn and Pb isotopes. \label{fig:gap}
  }
\end{figure}

In figure \ref{fig:gap}, we show typical examples of the neutron pairing gap and OES for Sn and Pb isotopes.
The experimental data are evaluated with equation~(\ref{eq:OES2}) using the values in \cite{ame2012}, while the average pairing gap (\ref{eq:gaprho}) is computed
using the UNEDF1-HFB EDF \cite{0954-3899-42-3-034024} with three kinds of density-dependent 
pairing functionals. The pairing strengths are fitted to reproduce the neutron OES at $^{120}$Sn. The general trends agree quite well, but the pairing gap does not have sensitivity to determine
the density dependence of the pairing functional through the OES.

It has been pointed out many times that the pairing gap is not an appropriate quantity to estimate the OES.
Existence of the multiple definitions of the pairing gaps such as equations (\ref{eq:gaprho}) and (\ref{eq:gapkap}), and also in the experimental OES (such as four-point, five-point formulae) makes the unique comparison impossible.
In order to avoid the above issue, one has to evaluate the OES directly by computing the
odd-mass systems.
As the time-reversal symmetry is broken in the odd-mass systems, 
the precise computation of the odd-mass systems is still a challenging problem in the nuclear DFT \cite{PhysRevC.81.024316}.
In addition to the blocking prescription, the time-odd functionals are activated, and their coupling constants are less known than those of the time-even functionals \cite{PhysRevC.93.014304}.

\subsection{Pairing rotational moment of inertia}

Another important aspect of the pairing correlation is the spontaneous breaking of the gauge symmetry.
The mean-field solution with the pairing correlations becomes superconducting, and is no longer the eigenstate of the particle number operator.
There appears a new collective degree of freedom in the symmetry-broken state, called 
zero-energy NG mode \cite{PhysRev.112.1900,PhysRev.117.648,INC_19_154}.
The pairing NG mode, generated from the broken U(1) gauge symmetry is called pairing rotation that corresponds to the rotation of a complex phase \cite{Brink-Broglia,Broglia20001}.
The pairing rotation is a zero-energy excitation in the intrinsic superconducting frame, while
in the laboratory frame where the particle number is conserved,
the ground states of the even-even nuclei can be regarded as a pairing rotational band.

For a system in which the proton shell is closed and the neutron shell is open and superconducting, the ground-state energy for a system with $N$ neutrons can be
described from a reference system with $N_0$ neutrons by expanding the energy
up to the second order with respect to the particle number fluctuation
$N- N_0$,
\begin{align}
  E(N) = E(N_0) + \lambda_n(N_0) (N - N_0) + \frac{1}{2{\cal J}_n(N_0)} (N - N_0)^2,
\end{align}
where $\lambda_n(N_0) = {\rm d}E/{\rm d}N|_{N=N_0}$ is the neutron chemical potential,
and ${\cal J}_n(N_0) = {\rm d}^2E/{\rm d}N^2|_{N=N_0}^{-1}$ is the neutron pairing rotational moment of inertia.
The second order term in $N-N_0$ is the pairing rotational energy.

Experimentally, the neutron chemical potential $\lambda_n(N)$ can be extracted from the two-neutron separation energy
\begin{align}
  \lambda_n(N) = - \frac{1}{4}[ S_{2n}(N+2) + S_{2n}(N)],
\end{align}
while the neutron pairing rotational moment of inertia ${\cal J}_n(N)$ can be evaluated by the inverse of double binding-energy differences.
\begin{align}
  {\cal J}_{n}(N) = \frac{4}{ E(N + 2)  - 2E(N) + E(N - 2)} = 
    \frac{4}{ S_{2n}(N) - S_{2n}(N+2)}. \label{eq:DBED}
\end{align}
We have proposed that the pairing rotational moment of inertia
is an excellent pairing observable \cite{PhysRevLett.116.152502}.
The moment of inertia of the pairing rotation is a quantity related to the gauge symmetry breaking due to the pairing correlations.
The inertia is evaluated using the ground-state energies of even-$N$ systems only, and thus there is no ambiguity from the unknown time-odd functionals or the approximations when computing the odd-mass systems. We note the neutron pairing rotational moment of inertia is proportional to 
the inverse of the two-neutron shell gap indicator $\delta_{2n}$.

\section{Finite-amplitude method (FAM) for NG modes \label{sec:FAM}}

The pairing rotational moment of inertia is originally defined as a second derivative of the energy
with respect to the particle number.
Theoretically it can be evaluated from three HFB calculations using equation~(\ref{eq:DBED}), but
this expression is regarded as an approximation to the second derivative.
The quasiparticle random-phase approximation (QRPA) allows us to derive the pairing rotational moment of inertia as an exact second derivative (Thouless-Valatin moment of inertia \cite{Thouless1962211}).
The zero-energy NG modes are the solutions of the QRPA equations.
The QRPA equations in the $PQ$ (momentum-coordinate) representation
are written as \cite{Ring-Schuck}
\begin{align}
  \begin{pmatrix} A & B \\ B^\ast & A^\ast \end{pmatrix} \begin{pmatrix} P_i \\ -P_i^\ast \end{pmatrix}
  =& {\rm i}\Omega_i^2 {\cal J}_i \begin{pmatrix} Q_i \\ Q_i^\ast \end{pmatrix}, \\
\begin{pmatrix} A & B \\ B^\ast & A^\ast \end{pmatrix} \begin{pmatrix} Q_i \\ -Q_i^\ast \end{pmatrix}  
  =&
  -\frac{{\rm i}}{{\cal J}_i}
   \begin{pmatrix} P_i \\ P_i^\ast \end{pmatrix},  
\end{align}
where $A$ and $B$ are QRPA matrices that include the residual interactions,
$Q_i$ and $P_i$ are the two-quasiparticle matrix elements of the coordinate and momentum operators
$\hat{Q}_i$ and $\hat{P}_i$
(normalized with
$\langle[\hat{Q}_i,\hat{P}_j]\rangle={\rm i}\delta_{ij}$) for the eigen mode with the frequency $\Omega_i$
and the inertia ${\cal J}_i$.
For the zero-energy neutron pairing rotational mode in proton-shell closed nuclei,
$\Omega_i=0$,
the momentum and coordinate operators become the neutron particle number operator $\hat{N}_n$
and its canonically conjugate neutron gauge angle operator $\hat{\Theta}_n$, respectively.
The Thouless-Valatin moment of inertia for the neutron pairing rotational mode is 
given by
\begin{align}
  {\cal J}_{n} = 2 N^{20}_{n} (A+B)^{-1} N^{20}_{n},  \label{eq:TVMOI}
\end{align}
where $N^{20}_{n}$ is the two-quasiparticle matrix elements of the $\hat{N}_{n}$ operator (assumed to be real).

Because the dimensions of the two-quasiparticles (the dimensions of the $A$ and $B$ matrices) are very large, and additional model space truncation is necessary for calculating
the Thouless-Valatin inertia directly from 
equation (\ref{eq:TVMOI}).
Instead, we evaluate the Thouless-Valatin inertia in an equivalent alternative approach,
the FAM for the linear response theory of nuclear DFT \cite{nakatsukasa:024318,PhysRevC.84.014314}.

The FAM is derived from the small-amplitude limit of the time-dependent HFB (TDHFB) equation,
thus it is formally equivalent to the QRPA.
The TDHFB equation is written as
\begin{align}
  {\rm i} \frac{\partial}{\partial t} a_\mu(t) = [ \hat{H}(t) + \hat{F}(t), a_\mu(t)]
  \label{eq:TDHFB}
\end{align}
with a weak external field that depends on time
\begin{align}
  \hat{F}(t) =& \eta(\hat{F} e^{-{\rm i}\omega t} + \hat{F}^\dag e^{{\rm i}\omega t}), \\
  \hat{F} =& \sum_{\mu\nu}\left[  F^{20}_{\mu\nu} a^\dag_\mu a^\dag_\nu + F^{02}_{\mu\nu} a_\nu a_\mu \right].
\end{align}
This introduces the oscillations with the same frequencies for
the quasiparticle wave functions, and the mean-field Hamiltonian.
The oscillating part of
the quasiparticle $\delta a_\mu(t) = a_\mu(t)e^{-{\rm i}E_\mu t}  - a_\mu $
and
the Hamiltonian $\delta \hat{H}(t) = \hat{H}(t) - \hat{H}_{\rm HFB}$
are expressed as
\begin{align}
      \delta a_\mu(t) =& \eta \sum_\nu a_\nu^\dag [ X_{\nu\mu}(\omega) e^{-{\rm i}\omega t}
      + Y_{\nu\mu}^\ast (\omega) e^{{\rm i}\omega t}], \\
  \delta\hat{H}(t) =& \eta [ \delta H(\omega) e^{-{\rm i}\omega t} + \delta H(\omega)^\dag e^{{\rm i}\omega t}], \\
  \delta H(\omega) =& \frac{1}{2} \sum_{\mu\nu} [ \delta H^{20}_{\mu\nu}(\omega) a^\dag_\mu a^\dag_\nu + \delta H^{02}_{\nu\mu}(\omega) a_\nu a_\mu].
\end{align}
From the first order terms in the small parameter $\eta$ in the TDHFB equation (\ref{eq:TDHFB}), we have the linear response equations
\begin{align}
  (E_\mu + E_\nu - \omega) X_{\mu\nu}(\omega) + \delta H^{20}_{\mu\nu}(\omega) = -F^{20}_{\mu\nu}, \label{eq:FAM1}\\
  (E_\mu + E_\nu + \omega) Y_{\mu\nu}(\omega) + \delta H^{02}_{\mu\nu}(\omega) = -F^{02}_{\mu\nu}, \label{eq:FAM2}
\end{align}
or equivalently
\begin{align}
  \begin{pmatrix} X(\omega) \\ Y(\omega)  \end{pmatrix}
  =
  - \left[ \begin{pmatrix} A & B \\ B^\ast & A^\ast \end{pmatrix}
    -
    \omega \begin{pmatrix} 1 & 0 \\ 0 & -1 \end{pmatrix}
    \right]^{-1}
  \begin{pmatrix} F^{20} \\ F^{02} \end{pmatrix}. \label{eq:linearresponse}
\end{align}

The FAM enables us to evaluate the fluctuations of the Hamiltonian, $\delta H^{20}$ and $\delta H^{02}$, very efficiently.
These quantities are written in terms of $A$ and $B$ matrices and QRPA wave functions $X$ and $Y$ (see equation~(9) in~\cite{PhysRevC.87.064309}). In the FAM, they are evaluated as follows.
The fluctuations, $\delta H^{20}$ and $\delta H^{02}$, together with other one-quasiparticle-one-quasihole part $\delta H^{11}$ are
connected to the particle-hole and particle-particle fluctuations by a Bogoliubov transformation
\begin{align}
  \begin{pmatrix}
    \delta H^{11} & \delta H^{20} \\ -\delta H^{02} & -(\delta H^{11})^T
  \end{pmatrix}
  =
  \begin{pmatrix} U^\dag & V^\dag \\ V^T & U^T \end{pmatrix}
  \begin{pmatrix} \delta h & \delta \tilde{h}^{(+)} \\ - \delta\tilde{h}^{(-)\ast} & - \delta h^T \end{pmatrix}
  \begin{pmatrix} U  & V^\ast \\ V & U^\ast \end{pmatrix},
\end{align}
where $U$ and $V$ are the HFB wave functions.
Then the fluctuations of the particle-hole and pairing parts of the Hamiltonian matrix are written in terms of the wave functions
\begin{align}
  \delta h =& \frac{ h[ U_a^\ast,V^\ast_a, U_b,V_b] - h[U^\ast,V^\ast, U,V] }{\eta},  \label{eq:dh}\\
  \delta \tilde{h}^{(+)} =& \frac{ \tilde{h}[ U_a^\ast,V^\ast_a, U_b,V_b] - \tilde{h}[U^\ast,V^\ast, U,V] }{\eta}, \label{eq:dAD}\\
  \delta \tilde{h}^{(-)} =& \frac{ \tilde{h}[ U_b^\ast,V^\ast_b, U_a,V_a] - \tilde{h}[U^\ast,V^\ast, U,V] }{\eta}, \label{eq:dBD}
\end{align}
with
\begin{align}
  U_a =& U + \eta V^\ast X^\ast, &
  V_a =& V + \eta U^\ast X^\ast, \\
  U_b =& U + \eta V^\ast Y, &
  V_b =& V + \eta U^\ast Y,
\end{align}
where $h[U^\ast,V^\ast,U,V]$ and $\tilde{h}[U^\ast,V^\ast,U,V]$ are the particle-hole and pairing part of the Hamiltonian
as functions of wave functions. From equations (\ref{eq:dh})-(\ref{eq:dBD}) the fluctuations of the
particle-hole and pairing parts can be evaluated essentially using the routine that computes $h$ and $\tilde{h}$ from the HFB wave function, and the functional derivative is replaced by a numerical derivative with a finite-amplitude small parameter $\eta$.
This is the reason the method is called finite-amplitude method.
Using the technique above, we can solve equations (\ref{eq:FAM1}) and (\ref{eq:FAM2}) iteratively.

The output of the FAM is the strength functions.
The expression of the strength function in the $PQ$ representation is given by \cite{Blaizot-Ripka,PhysRevC.92.034321}
\begin{align}
  S(\hat{F},\omega) =& \sum_{\mu<\nu} F^{20\ast}_{\mu\nu} X_{\mu\nu}(\omega) + F^{02\ast}_{\mu\nu}
  Y_{\mu\nu}(\omega) \nonumber \\
  =& \sum_i \frac{1}{\omega^2 - \Omega_i^2}
  \left\{
  \frac{1}{{\cal J}_i} |\langle P_i|\hat{F}|0\rangle|^2 + {\cal J}_i \Omega_i^2 |\langle Q_i|\hat{F}|0\rangle|^2 \right. \nonumber \\ & \left. 
  + {\rm i}\omega \left( \langle Q_i|\hat{F}|0\rangle^\ast \langle P_i|\hat{F}|0\rangle - \langle P_i|\hat{F}|0\rangle^\ast
  \langle Q_i|\hat{F}|0\rangle \right)
  \right\}, \label{eq:SFomega}
\end{align}
where
$\langle P_i|\hat{F}|0\rangle= \langle [\hat{P}_i,\hat{F}]\rangle$
and
$\langle Q_i|\hat{F}|0\rangle= \langle [\hat{Q}_i,\hat{F}]\rangle$.
In the standard calculations using the FAM, the imaginary part of the strength function $-{\rm Im}S(\hat{F},\omega_\gamma)/\pi$ for a complex frequency $\omega_\gamma=\omega + {\rm i}\gamma$ provides the
Lorentzian-smeared strength distribution.

The expression above is not valid for $\omega=0$ when the NG mode is present ($\Omega_{\rm NG}=0$). The strength function for the momentum operator of the NG mode is computed directly from equation~(\ref{eq:linearresponse}) 
\begin{align}
  S(\hat{N}_{n},\omega=0) =& \sum_{\mu<\nu} (N^{20}_n)_{\mu\nu} X_{\mu\nu}(0) + (N^{20}_n)_{\mu\nu} Y_{\mu\nu}(0) \nonumber \\
  =& - 2N^{20}_{n}(A+B)^{-1}N^{20}_{n} = - {\cal J}_n. \label{eq:FAM-PRMOI}
\end{align}
Therefore in the FAM, the Thouless-Valatin moment of inertia is given from the strength function for the zero-energy response with the broken-symmetry operator.
This expression has been checked numerically for the center-of-mass motion and the pairing rotations \cite{PhysRevC.92.034321}.

\begin{figure}
  \includegraphics[width=160mm]{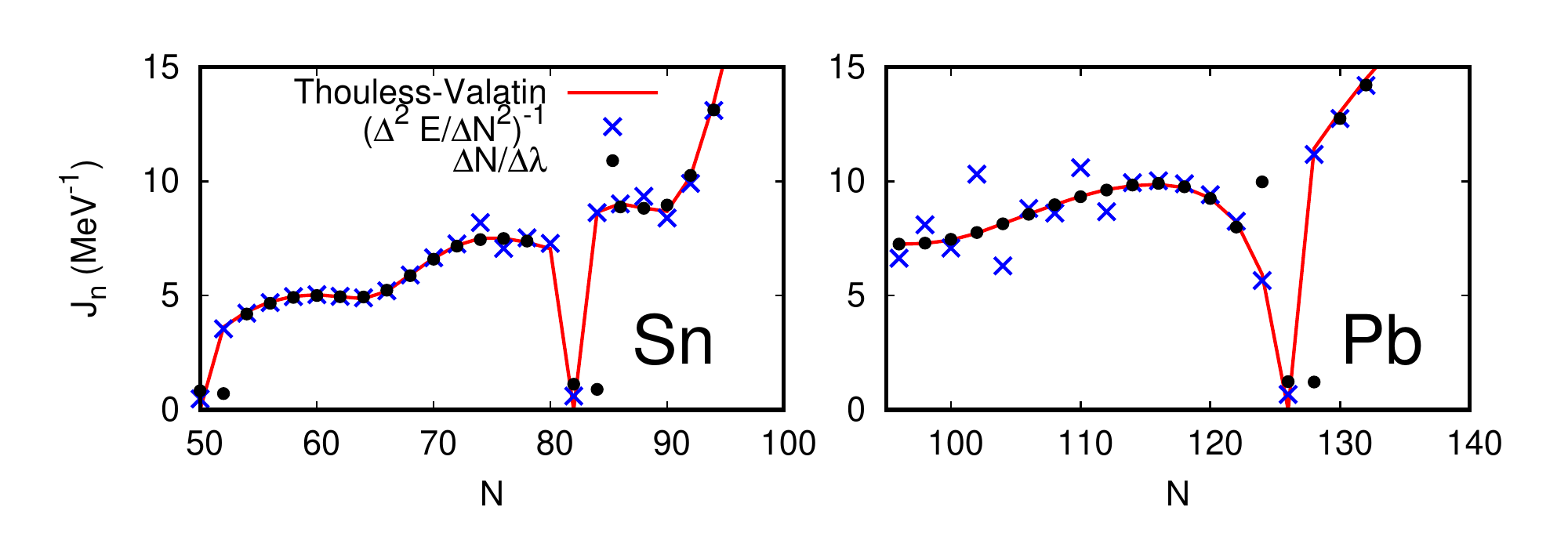}
  \caption{ The neutron pairing rotational moment of inertia of Sn and Pb isotopes computed as Thouless-Valatin inertia using the FAM (line), inverse of the double binding-energy differences using three HFB states (cross), and finite difference of the chemical potential using two HFB states (dot). $\Delta N=2$ is used in the finite difference evaluations.
    \label{fig:comparison}
   }  
\end{figure}

We have compared the neutron pairing rotational moment of inertia evaluated from the Thouless-Valatin inertia using equation (\ref{eq:FAM-PRMOI}) and from the inverse of the double binding-energy differences (\ref{eq:DBED}) using three HFB states in figure \ref{fig:comparison} in order to see the consistency between these two evaluations.
The calculations are performed for Sn and Pb isotopes, and the pairing functional of the volume type is employed with the parameters described in the next section.
The agreement between the double binding energy differences and the Thouless-Valatin inertia is generally good,
but in some nuclei there is a discrepancy.
We also evaluate the pairing rotational moments of inertia using the same HFB states but with the relation between the particle number and chemical potential
${\cal J}_n(N) = [ {\rm d}\lambda_n(N)/{\rm d}N]^{-1} \approx (2\Delta N) / [ \lambda_n(N+\Delta N) - \lambda_n(N-\Delta N)]$.
The pairing rotational moment of inertia derived from the chemical potential agrees with the
Thouless-Valatin inertia much better than that from the double binding-energy differences of the three HFB states does,
because in finite difference the numerical error is generally larger in evaluating the second derivative than
in evaluating the first derivative.
The total energy is affected more by the fluctuation around the mean field than the chemical potential is.
However next to the closed shell, the evaluation from the chemical potential deviates completely because the chemical potential at the closed shell is not uniquely defined (in this calculation the energy of the last occupied orbit is used as the chemical potential
when the pairing collapses).

\section{Pairing EDF \label{sec:ppEDF}}

Skyrme EDF is a zero-range energy density functional,
widely used for the nuclear EDF calculations.
It is originally based on the Skyrme effective interaction \cite{SKYRME1958615,PhysRevC.5.626}.
Therefore the coupling constants of the Skyrme EDF is derived from the parameters of the Skyrme effective interactions.
In the concept of the generalized EDF, however, the EDF can be parametrized independent of the effective interaction.
The bilinear density form together with the density-dependent term is assumed in the particle-hole Skyrme EDF,
and the coupling constants are optimized using a large number of experimental data sets.
Recent effort in this direction produces a series of the Skyrme-EDF parameter sets:
SV$_{\rm min}$ \cite{PhysRevC.79.034310}, 
UNEDF0 \cite{PhysRevC.82.024313}, UNEDF1 \cite{PhysRevC.85.024304},
UNEDF1-HFB \cite{0954-3899-42-3-034024}, and UNEDF2 \cite{PhysRevC.89.054314}.

In the Skyrme EDF,
a simple zero-range pairing force with density dependence is commonly used \cite{RevModPhys.75.121}
\begin{align}
  \tilde{\chi}_t(\bm{r}) = \frac{V_t}{4} \left[ 1 - \eta \left(\frac{\rho_0(\bm{r})}{\rho_c}\right)^\beta\right] |\tilde{\rho}_t(\bm{r})|^2,
  \label{eq:pairing}
\end{align}
where $\tilde{\rho}_t(\bm{r})$ is a pair density of neutron or proton,
$\rho_0(\bm{r}) = \rho_n(\bm{r}) + \rho_p(\bm{r})$ is the isoscalar particle-hole density, $\rho_c=0.16$ fm$^{-3}$, and
$V_t$ is the strength parameter.
The parameter $\beta$ is usually taken to be one.
The parameter $\eta$ controls the density dependence of the pairing functional;
$\eta=0$ for volume pairing without density dependence, $\eta=1$ for surface pairing, and
$\eta=0.5$ for mixed pairing.
The density dependence of the pairing interaction is not fully determined yet,
and is an open problem.
In the recent optimized Skyrme functionals, values around the mixed pairing is adopted
\cite{PhysRevC.79.034310,PhysRevC.89.054314}.

There are many attempts to extend the pairing functional from the standard one (\ref{eq:pairing}).
The isoscalar and isovector density dependence are proposed and investigated to improve the behavior of the pairing in the nuclear matter \cite{PhysRevC.76.064316} and neutron-rich nuclei \cite{PhysRevC.77.054309, PhysRevC.77.064319,PhysRevC.80.064301,PhysRevC.86.034333}.
Pairing functionals of the Fayans model \cite{FAYANS199619,Fayans200049} with density gradient terms are recently discussed \cite{PhysRevC.95.064328}.

In this paper, we extend the pairing EDF in a straightforward way.
The form of the pairing EDF in equation~(\ref{eq:pairing}) corresponds to the $t_0$ and $t_3$ terms in the
original Skyrme effective interaction.
More terms are derived to the pairing EDF from the original Skyrme effective interaction.
In the Skyrme SkP parametrization \cite{Dobaczewski1984103}, both the particle-hole and pairing parts
are derived from a single Skyrme effective interaction.
Although the pairing interaction derived from the Skyrme force is known to be unrealistic, the concept of the generalized DFT allows us to fit the pairing coupling constants independently.

Within the bilinear form of the pair density (with isoscalar particle-hole density dependence), the most general form of the isovector (nn and pp) pairing functional is given by \cite{PhysRevC.69.014316}
\begin{align}
  \tilde{\chi}_t(\bm{r}) =& \tilde{C}_t^{\rho}[\rho_0] |\tilde{\rho}_t|^2
  + \tilde{C}^{\Delta\rho}_t {\rm Re}( \tilde{\rho}_t^\ast \Delta \tilde{\rho}_t)
  + \tilde{C}^\tau_t {\rm Re}(\tilde{\rho}_t^\ast \tilde{\tau}_t) \nonumber \\
  &+ \tilde{C}^{J0}_t |\tilde{J}_t|^2
  + \tilde{C}^{J1}_t |\tilde{\bm{J}}_t|^2
  + \tilde{C}^{J2}_t |\tilde{\underline{\sf J}}_t|^2
  + \tilde{C}^{\bm{\nabla} J}_t {\rm Re}( \tilde{\rho}_t^\ast \bm{\nabla}\cdot \tilde{\bm{J}}_t), \label{eq:ppgenEDF}
\end{align}
where independent coupling constants are assumed for neutrons and protons,
and the local pair densities are given from the non-local pair densities $\tilde{\rho}_t(\bm{r},\bm{r}')$ and $\tilde{\bm{s}}_t(\bm{r},\bm{r}')$ as 
\begin{align}
  \tilde{\rho}_t(\bm{r}) =& \tilde{\rho}_t(\bm{r},\bm{r}), \\
  \tilde{\tau}_t(\bm{r}) =& (\bm{\nabla}\cdot\bm{\nabla}')\tilde{\rho}_t(\bm{r},\bm{r}')\Big|_{\bm{r}=\bm{r}'}, \\
  \tilde{\sf J}_{tab}(\bm{r}) =& \frac{1}{2{\rm i}} [ (\nabla_a - \nabla'_a)  \tilde{s}_{tb}(\bm{r},\bm{r}')]\Big|_{\bm{r}=\bm{r}'},\\
  \tilde{J}_t(\bm{r}) =& \sum_{a} \tilde{\sf J}_{taa}(\bm{r}), \\
  \tilde{\bm{J}}_{ta}(\bm{r}) =& \sum_{bc} \varepsilon_{abc} \tilde{\sf J}_{tbc}(\bm{r}), \\
  \tilde{\underline{\sf J}}_{tab}(\bm{r}) =& \tilde{\sf J}_{tab}(\bm{r}) + \tilde{\sf J}_{tba}(\bm{r}) - \frac{1}{3}\tilde{J}_t(\bm{r}).
\end{align}
In this paper, we reduce the number of degrees of freedom to make the discussion simpler.
We first assume the local gauge invariance \cite{PhysRevC.69.014316,Engel1975215,PhysRevC.52.1827}, which is the symmetry associated with the local gauge transformation of the wave function
\begin{align}
  |\Psi'\rangle  = e^{{\rm i}\phi(\bm{r})}|\Psi\rangle.
\end{align}
The local gauge invariance of the pairing EDF provides two conditions to the coupling constants
\begin{align}
  \tilde{C}^{\Delta\rho}_t = -\frac{1}{4} \tilde{C}^\tau_t, \quad
  \tilde{C}^{\bm{\nabla} J}_t = 0.
\end{align}
We do not consider the tensor pairing functional in this paper for simplicity ($\tilde{C}^{J0}_t=\tilde{C}^{J1}_t=\tilde{C}^{J2}_t=0$).
Finally we have a pairing EDF of the following form
\begin{align}
  \tilde{\chi}_t(\bm{r}) = \tilde{C}_t^\rho [\rho_0]|\tilde{\rho}_t|^2 + \tilde{C}_t^{\Delta\rho}[ {\rm Re}(\tilde{\rho}_t^\ast \Delta
    \tilde{\rho}_t) - 4 {\rm Re} ( \tilde{\rho}_t^\ast \tilde{\tau}_t) ]. \label{eq:ppEDF}
\end{align}
The second term originally corresponds to the momentum-dependent terms with $t_1$ parameter in the Skyrme effective interaction.

The pairing potential for the pairing EDF (\ref{eq:ppEDF}) is given by
\begin{align}
  \tilde{h}_t(\bm{r}) =& \tilde{U}_t(\bm{r}) - \bm{\nabla}\cdot \tilde{M}_t(\bm{r})\bm{\nabla}, \\
  \tilde{U}_t(\bm{r}) =& 2 \tilde{C}_t^\rho[\rho_0] \tilde{\rho}_t + 2 \tilde{C}_t^{\Delta\rho}\Delta\tilde{\rho}_t
  + \tilde{C}_t^\tau \tilde{\tau}_t, \\
  \tilde{M}_t(\bm{r}) =&  \tilde{C}_t^\tau \tilde{\rho}_t.
\end{align}

\section{Pairing rotational moments of inertia in Sn and Pb \label{sec:calc}}

We discuss the effect of the pair-density derivative terms introduced in the previous section on the pairing rotational moments of inertia.
As typical examples of the pairing rotations,
we focus on the neutron pairing rotations in Sn and Pb isotopes.

The calculation for the DFT and FAM for the Thouless-Valatin moments of inertia
is performed using the HFBTHO code \cite{Stoitsov200543,Stoitsov20131592}
and its extension to the FAM with the module developed in \cite{PhysRevC.84.041305,PhysRevC.92.034321}.
Twenty harmonic oscillator major shells together with 60 MeV pairing cutoff energy is used.
The parameters for the Gauss quadratures are $N_{\rm GH}=N_{\rm GL}=40$, and $N_{\rm Leg}=80$.
The UNEDF1-HFB functional \cite{0954-3899-42-3-034024} is used for the particle-hole part of the EDF.
In the pairing functional~(\ref{eq:ppEDF}), there are two pairing coupling constants.
In order to see the dependence on the pair-density derivative terms,
we use several sets of the coupling constant $\tilde{C}^{\Delta\rho}_n=0, -20, -40, -60$, and $-80$ MeV fm$^{5}$.
For each value of $\tilde{C}^{\Delta\rho}_n$, we fit the pairing strength $V_n$
using the experimental value of the neutron pairing rotational moment of inertia at $^{120}$Sn (6.64 MeV$^{-1}$).
Three types of density dependence, volume, mixed, and surface are considered.
The coupling constants used in this analysis are summarized in table \ref{table:parameters}.

We analyze the effect of the pair-density derivative coupling constants
on the basic quantities, rms radius and HFB energies in $^{120}$Sn and $^{198}$Pb.
For the pairing functionals listed in table \ref{table:parameters}, 
the rms radius agrees within 0.03 fm and 0.02 fm for $^{120}$Sn and $^{198}$Pb, that is about 0.6\% and 0.4\%, respectively,
and the HFB energy agrees within 1.7 MeV and 5.0 MeV that is 0.2\% and 0.3\%, respectively.
The effect of the pair-density derivative coupling constants on these quantities are similar to the effect from different density dependence of the pairing functional. Therefore we focus on the pairing properties in the following.

\begin{table}
  \caption{Pairing coupling constants $V_n$ fitted to the neutron pairing rotational moment of inertia in $^{120}$Sn for each value of $\eta$ and $\tilde{C}_n^{\Delta\rho}$.}
  \footnotesize
  \label{table:parameters}
  \begin{center}
  \begin{tabular}{@{}ccccc} \br
    $\tilde{C}_n^{\Delta\rho}$ (MeV fm$^{5}$) & $\tilde{C}^{\tau}_n$ (MeV fm$^{5}$)&
     \multicolumn{3}{c}{$V_n$ (MeV fm$^{3}$)} \\ \cline{3-5}
    & &    $\eta=0$ & $\eta=0.5$ & $\eta=1$
    \\ \mr
\,\,\,\,\,\,\,\,0 &  \,\,\,\,\,\,0    & $-$148.6 & $-$225  & $-$389\\
$-$20 & \,\,\,80  & $-$593\,\,\,\,   & $-$820  & $-$1093\,\,\,\\
$-$40 & 160 & $-$774\,\,\,\,   & $-$1072\,\,\, & $-$1450\,\,\,\\
$-$60 & 240 & $-$720\,\,\,\,   &  $-$993 & $-$1499\,\,\,\\
$-$80 & 320 & $-$354\,\,\,\,   &  $-$521 & $-$924 \\ \br
  \end{tabular}
  \end{center}
\end{table}
\normalsize

\begin{figure}[htbp]
  \includegraphics[width=160mm]{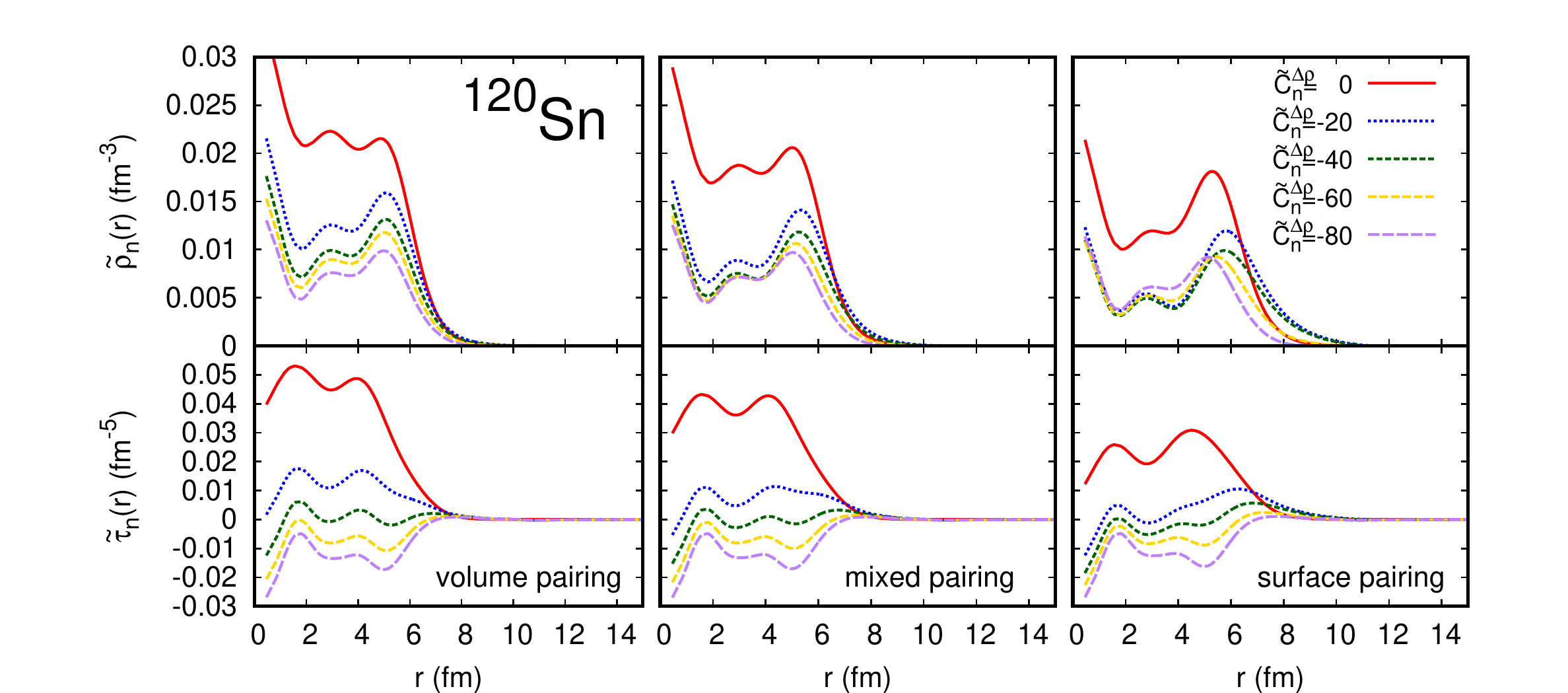}
  \caption{Neutron pair density $\tilde{\rho}_n(\bm{r})$ (upper) and neutron pair kinetic density
    $\tilde{\tau}_n(\bm{r})$ (lower) in $^{120}$Sn.
    The volume (left), mixed (middle) and surface (right)-type density-dependent pairing are used together with the pair-density derivative coupling constants $\tilde{C}_n^{\Delta\rho} = -4\tilde{C}_n^\tau$.
    \label{fig:density}}
\end{figure}

We begin with the effect of the pair-density derivative terms in $^{120}$Sn.
Figure~\ref{fig:density} shows the neutron pair-density distribution.
In the comparison of the isoscalar density dependence of the pairing EDF,
the pair density is rather constant inside the nucleus
when the volume pairing is employed, while it has a peak in the
nuclear surface region with the surface pairing.
The pair-density derivative terms in the EDF generally provide a peak in the surface region of the pair density.
This effect is similar for volume, mixed and surface pairing, and
with stronger pair-density derivative terms, almost similar pair-density distribution are obtained
for three kinds of isoscalar density dependence. The same trend is also seen in the pair kinetic density distribution. With the pair-density derivative terms, the sign of the pair kinetic density can change.

The pair-density distribution can be compared with figure 2 in \cite{PhysRevC.53.2809},
where the neutron pair density for $^{120}$Sn is calculated with full SkP interaction (pairing part is from SkP), and 
the particle-hole SkP interaction with the volume pairing (SkP$^\delta$).
The pair density obtained by UNEDF1-HFB with volume pairing without pair-density derivative terms is close to the
one obtained by SkP$^\delta$.
The pair density obtained with the full SkP interaction has a strong peak in the surface area.
Such a pair density is obtained either with the surface pairing, or with the pair-density derivative terms.
We note that in the full SkP results, isoscalar density dependence with $\beta=1/6$ and pair-density derivative terms are included, and moreover, pair tensor density terms are included.

\begin{figure}[htbp]
    \includegraphics[width=170mm]{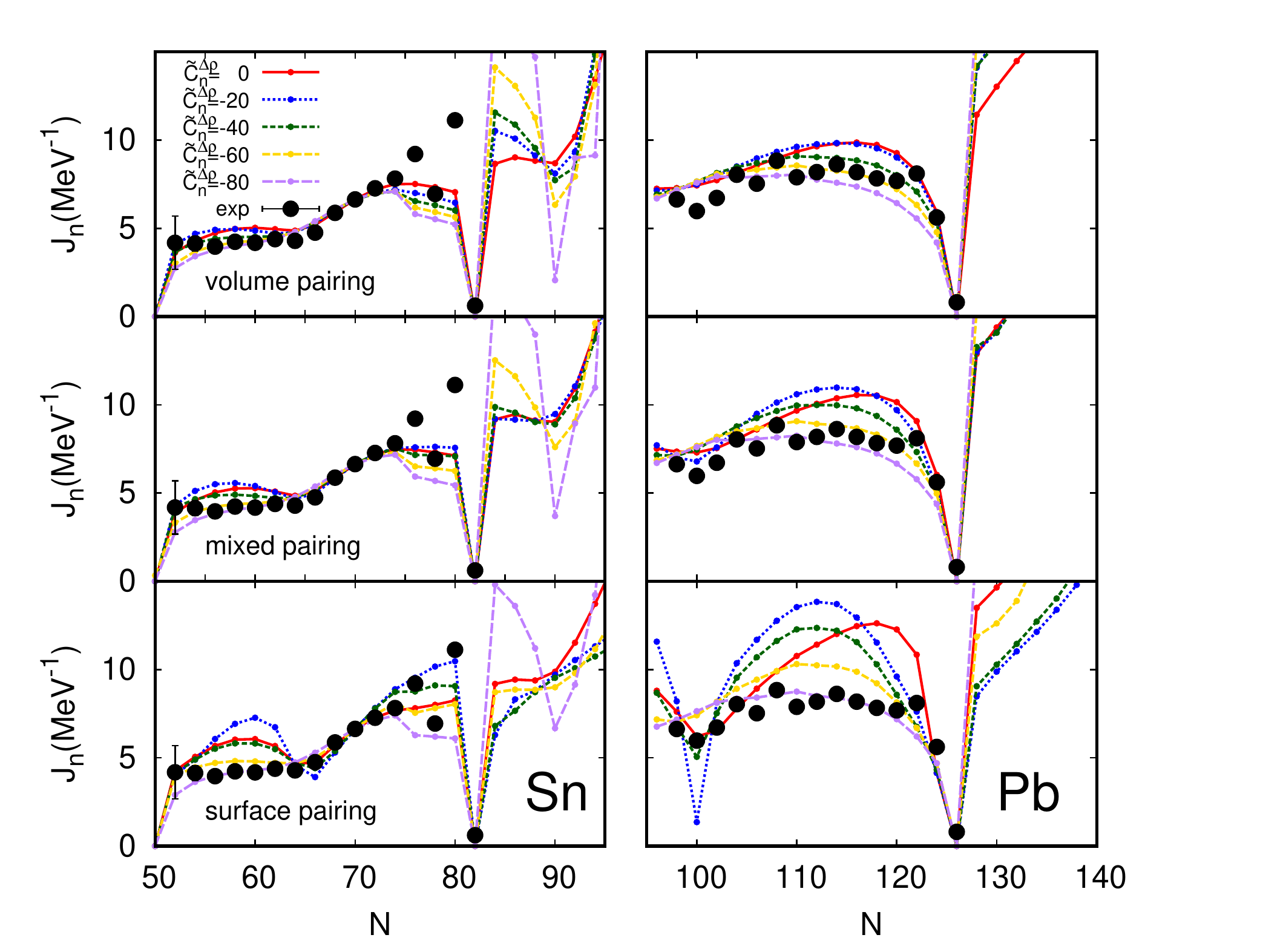}
  \caption{Neutron pairing rotational moment of inertia of Sn (left) and Pb (right) isotopes.
    The volume (top), mixed (middle), and surface (bottom) pairing are used together with the pair-density derivative terms.
    The experimental values are derived from the double binding-energy differences (\ref{eq:DBED}) using the values in \cite{ame2012}.
    \label{fig:TVMOI}}
\end{figure}

In figure \ref{fig:TVMOI} the neutron pairing rotational moment of inertia for Sn and Pb 
are shown.
The calculations without pair-density derivative terms ($\tilde{C}_n^{\Delta\rho}=0$) are comparable with the
previous results in \cite{PhysRevLett.116.152502}, where the same pairing functionals are used, but
the coupling constants are fitted to the neutron pairing gap in $^{120}$Sn.
The results of the volume and mixed pairing are almost the same as the previous one, because the coupling constants listed in Table \ref{table:parameters} are similar to the previous values,
while the surface pairing strength fitted to the neutron pairing rotational moment of inertia is smaller than the previous value
fitted to the neutron pairing gap ($V_n=-474.32$ MeV fm$^3$), and
the neutron pairing rotational moment of inertia is systematically larger than the values obtained in ~\cite{PhysRevLett.116.152502}.
In all cases, the results without the pair-density derivative terms cannot reproduce the trends in Sn and Pb simultaneously.
Theoretical values of the inertia overestimate the experimental ones in the light Sn region around $N=60$
and in the Pb isotopes around $N=110$-$120$. Thus none of the density dependence can reproduce the experimental trend precisely.

The pair-density derivative terms improve this situation.
The main effect of the pair-density derivative terms
in the neutron pairing rotational moment of inertia in Sn and Pb isotopes is
the decrease near $N=60$ and $N=78$ in Sn isotopes, and $N=110$-$120$ in Pb isotopes.
Irrespective of the density dependence of the pairing functional,
the pair-density derivative terms improve the quantitative agreement of the neutron
pairing rotational moment of inertia of both Sn and Pb isotopes.

\begin{figure}[htbp]
  \includegraphics[width=170mm]{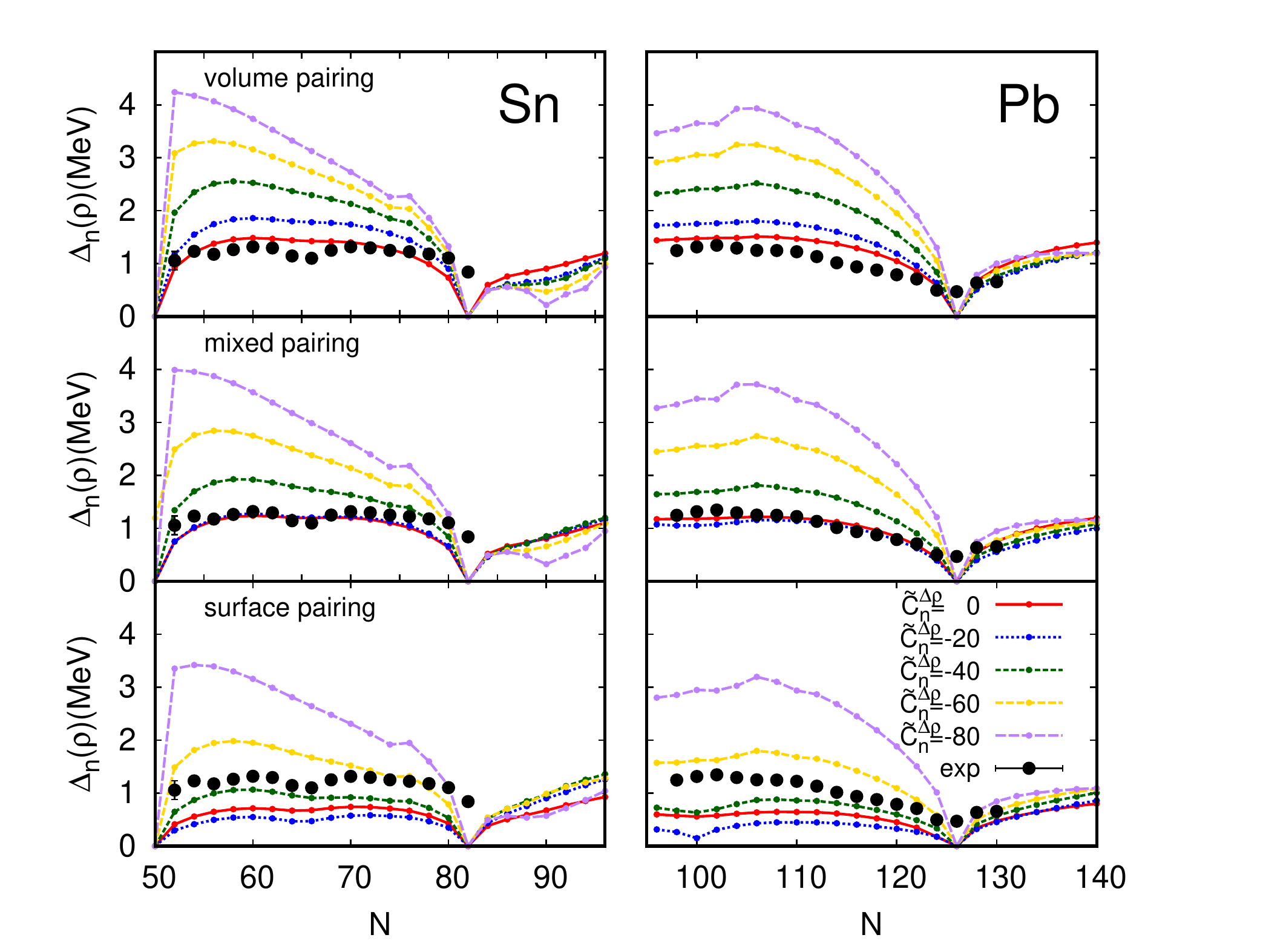}
     \caption{Neutron average pairing gap $\Delta_n(\rho)$ of Sn (left) and Pb (right) isotopes.
       The volume (top), mixed (middle), and surface (bottom) pairing are used together with the pair-density derivative terms.
       The experimental values are evaluated from equation (\ref{eq:OES2}) using the values in \cite{ame2012}.
       \label{fig:gaprho}}
\end{figure}

\begin{figure}[htbp]
    \includegraphics[width=170mm]{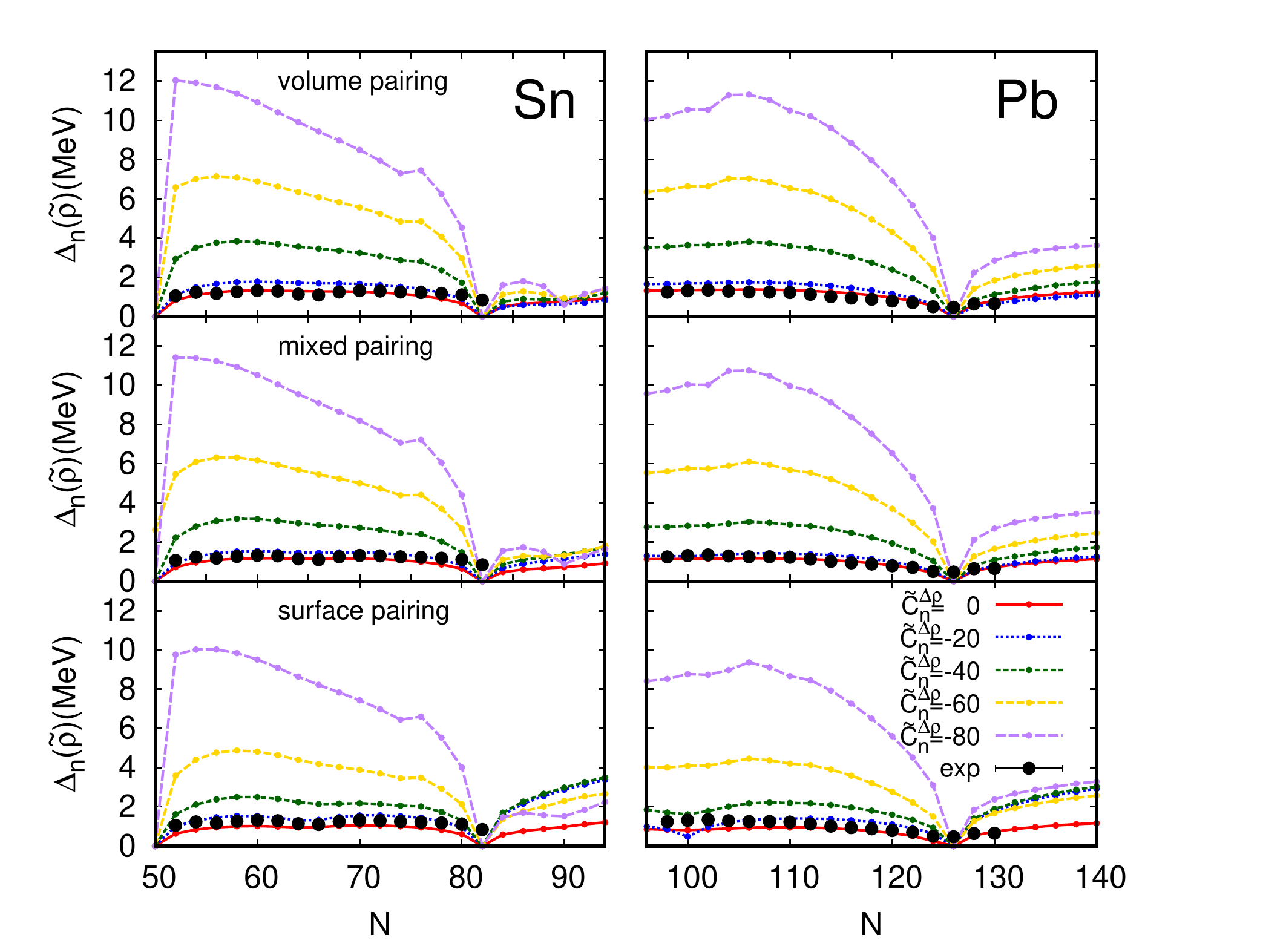}
    \caption{Same as figure \ref{fig:gaprho} but for the neutron average pairing gap $\Delta_n(\tilde{\rho})$. \label{fig:gapkap}}
\end{figure}

Next we analyze the pairing gaps.
We define two kinds of the average pairing gaps by extending
equations (\ref{eq:gaprho}) and (\ref{eq:gapkap}) as
\begin{align}
  \Delta_n(\rho) =& - \frac{ \Tr \tilde{h}_n \rho_n}{\Tr \rho_n} \nonumber \\
      =& - \frac{1}{N} \int {\rm d}\bm{r} \left\{ \left[      
      2 \tilde{C}_n^\rho[\rho_0]\tilde{\rho}_n(\bm{r}) + 2 \tilde{C}_n^{\Delta\rho} \Delta\tilde{\rho}_n(\bm{r})  + \tilde{C}_n^\tau \tilde{\tau}_n(\bm{r})
      \right] \rho_n(\bm{r}) + \tilde{C}^\tau_n \tilde{\rho}_n(\bm{r}) \tau_n(\bm{r})
    \right\}, \label{eq:gaprho2}
\end{align}
and
\begin{align}
  \Delta_n(\tilde{\rho}) =& \frac{ \Tr \tilde{h}_n \tilde{\rho}^\ast_n}{\Tr \tilde{\rho}_n} \nonumber \\
    =& - \frac{1}{\tilde{N}} \displaystyle\int {\rm d}\bm{r} \left\{
    2 \tilde{C}_n^\rho[\rho_0] |\tilde{\rho}_n(\bm{r})|^2 + 2 \tilde{C}_n^{\Delta\rho} {\rm Re} \left[ \tilde{\rho}_n^\ast(\bm{r}) \Delta\tilde{\rho}_n(\bm{r}) \right] + 2\tilde{C}_n^\tau {\rm Re} \left[ \tilde{\rho}_n^\ast(\bm{r}) \tilde{\tau}_n(\bm{r}) \right]
    \right\} \nonumber \\
    =& - \frac{2 E_n^{\rm pair} }{\tilde{N}}. \label{eq:gapkap2}
\end{align}
In the case of the present pairing EDF and choice of the phase (normally real representation is used for the pair density),
the pairing gap $\Delta_n(\tilde{\rho})$ is related to the total pairing energy.
Figures~\ref{fig:gaprho} and \ref{fig:gapkap} show the average pairing gap $\Delta_n(\rho)$ and $\Delta_n(\tilde{\rho})$, respectively. The pairing gap $\Delta_n(\rho)$ increases with the strength of the pair-density derivative term $\tilde{C}^{\Delta\rho}_n$.
The correspondence with the experimental OES is the best when pair-density derivative term is absent for the volume and mixed pairing, and at around $\tilde{C}_n^{\Delta\rho}=-50$ MeV fm$^{5}$ for the surface pairing.
The pairing gap $\Delta_n(\tilde{\rho})$ is more sensitive to the strength of the pair-density derivative term,
and it overestimates the experimental OES when $\tilde{C}^{\Delta\rho}_n$ is larger than $-20$ MeV fm$^5$.

Two problems arise from this analysis.
First, the sensitivity to the pair-density derivative term is different between
the two average pairing gaps $\Delta_n(\rho)$ and $\Delta_n(\tilde{\rho})$, thus it is not possible to determine this
coupling constant uniquely using the pairing gaps and experimental OES.
Second, in the neutron pairing rotational moment of inertia, the agreement with the experimental data is improved for large $\tilde{C}_n^{\Delta\rho}$ such as $-60$ and $-80$ MeV fm$^{5}$,
while the correspondence between the average pairing gap and OES is totally lost for large $\tilde{C}_n^{\Delta\rho}$.
This indicates that the conventional pairing gaps in the presence of the pair-density derivative term
do not correspond to the experimental OES due to the kinetic pairing contribution.
On the other hand, the pairing rotational moment of inertia is well defined with the
second derivative of the energy with respect to the particle number
(or double binding-energy difference of even-even nuclei), and is free from this problem.

\begin{figure}[htbp]
  \includegraphics[width=170mm]{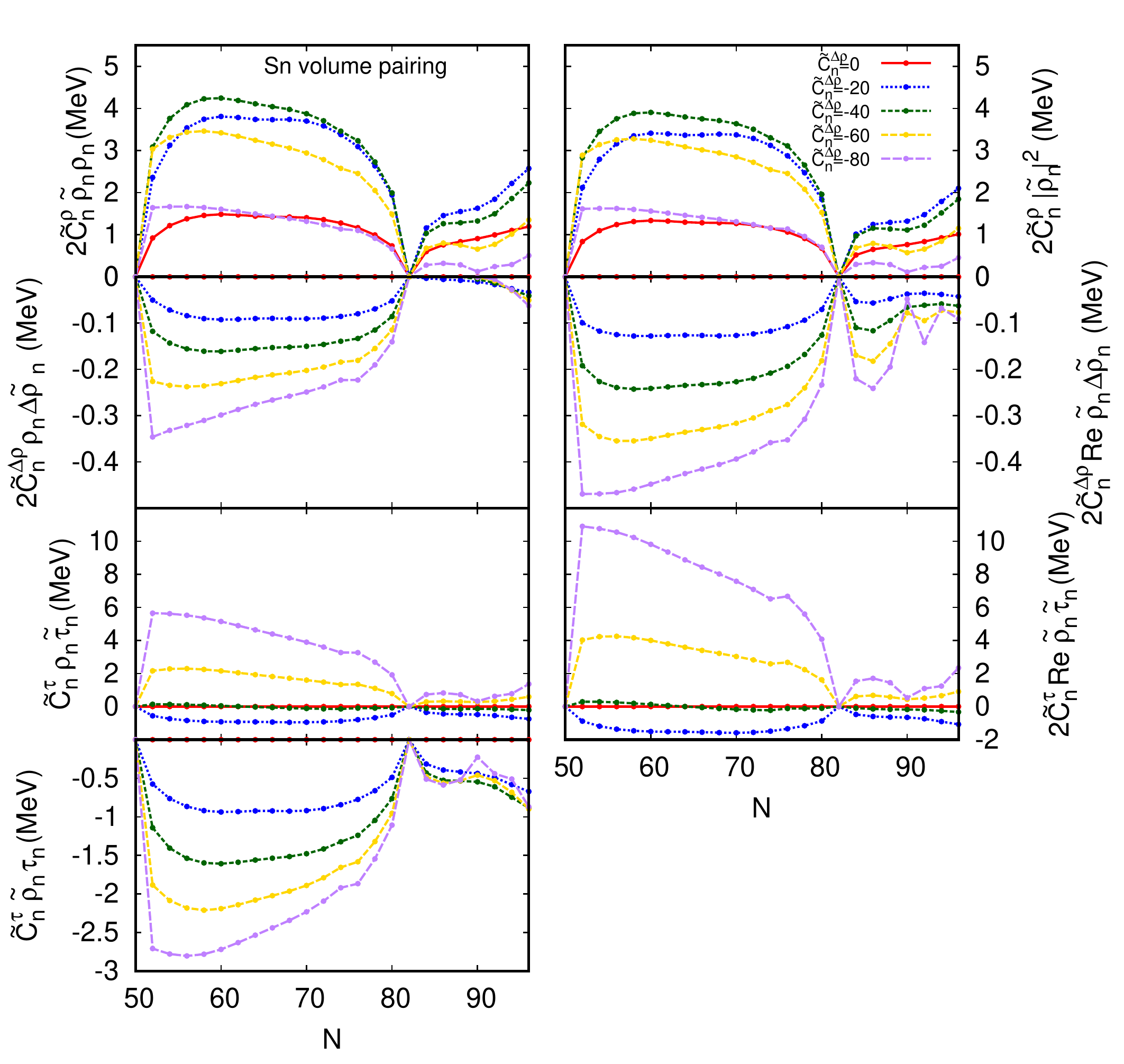}
  \caption{The contributions of each terms in equations~(\ref{eq:gaprho2}) and (\ref{eq:gapkap2}) to the pairing gap.
    Left four panels show the contributions from the four terms in equation (\ref{eq:gaprho2}), and right three panels show the contributions from the three terms in equation (\ref{eq:gapkap2}).
    \label{fig:gapterms}
  }
\end{figure}

As described in equations~(\ref{eq:gaprho2}) and (\ref{eq:gapkap2}), the pairing gaps are composed of several terms.
To analyze the origin of the differences between the two pairing gaps $\Delta(\rho)$ and $\Delta(\tilde{\rho})$, and deviation of the pairing gaps from the experimental OES, 
each term in the pairing gaps is plotted separately in figure \ref{fig:gapterms} in the case of the volume pairing.
The contribution to the pairing gap from the second term, which includes the pair density coupling with $\Delta \tilde{\rho}_n$
is one order smaller than the contributions from the other terms.
The conventional first term changes much with the pair-density derivative coupling constant $\tilde{C}^{\Delta\rho}_n$.
It increases and reaches its maximum at around $\tilde{C}^{\Delta\rho}_n=-40$ MeV fm$^5$, and then decreases for larger values of $\tilde{C}^{\Delta\rho}_n$. The increase of the pairing gap up to $\tilde{C}^{\Delta\rho}_n=-40$ MeV fm$^5$ is due to the increase of the first term.
Then for $\tilde{C}^{\Delta\rho}_n=-60$ MeV fm$^5$ and $-80$ MeV fm$^5$, the third term which includes the pair kinetic density $\tilde{\tau}_n$ suddenly increases, and provides the main contribution to the pairing gap. From the pair kinetic density for $^{120}$Sn in figure \ref{fig:density}, it can be related to the change of the sign in pair kinetic density.

In $\Delta(\tilde{\rho})$, the third term is proportional to $\tilde{C}^\tau_n$ with a factor of two,
while in $\Delta(\rho)$, the corresponding term is divided into the third and fourth terms.
The third terms in $\Delta(\rho)$ and $\Delta(\tilde{\rho})$ are very similar, except for the difference in the factor of two in $\Delta(\tilde{\rho})$.        
The fourth term that include the particle-hole kinetic density $\tau_n$ has opposite sign in $\Delta(\rho)$, and this
cancels the contribution from the third term partly in $\Delta(\rho)$.
We can say that the difference between the two pairing gaps are mainly from the differences between the particle-hole and pair kinetic densities $\tau_n[$ and $\tilde{\tau}_n$.

\section{Conclusions \label{sec:conclusions}}

We have extended the standard pairing EDF by including the pair-density derivative term that corresponds to the momentum-dependent term in the Skyrme effective interaction.
Assuming the local gauge invariance, one coupling constant is introduced
for neutrons and protons, respectively.
The pairing coupling constants are fitted to reproduce the pairing rotational moments of inertia,
which are microscopically computed as the Thouless-Valatin inertia using the finite-amplitude method for nuclear DFT. 
It is shown that the pair-density derivative term systematically improves the behavior of the neutron pairing rotational
moments of inertia in Sn and Pb isotopes.
The results indicate that the conventional theoretical average pairing gaps do not serve as measures of the
experimental odd-even mass staggering when the pair-density derivative term is present.

The results demonstrate that the pairing rotational moment of inertia is effective for constraining the pairing EDF including the pair-density derivative term.
Exploring the pairing tensor terms that are not taken into account in the present analysis is another interesting direction to investigate.
Future work should include the systematic and global optimization of the pairing EDF using the pairing rotational moments of inertia.

\section*{Acknowledgments}
Useful discussions with Witold Nazarewicz are gratefully acknowledged.
This work is supported by the JSPS KAKENHI Grand Numbers 16K17680 and 17H05194.
Numerical calculation was performed in the resources of High Performance Computing Center,
Institute for Cyber-Enables Research, Michigan State University,
and
the COMA (PACS-IX) System at the Center for Computational Sciences, University of Tsukuba, through the Interdisciplinary Computational Science Program 
and HPCI System Research Project (hp170144).

\section*{References}
\bibliographystyle{iopart-num}
\bibliography{prmoi}

\end{document}